\documentclass[aps,prl,reprint,amsmath,amssymb]{revtex4-1}
\usepackage{graphicx}
\usepackage{dcolumn}
\usepackage{epsfig}
\usepackage{bm}
\usepackage{amsmath}
\usepackage{amssymb}
\usepackage{float}
\def\be{\begin{eqnarray}}
\def\ee{\end{eqnarray}}

\def\={&=&}
\def\f{\frac}

\def\mc{\mathcal}

\def\be{\begin{eqnarray}}
\def\ee{\end{eqnarray}}

\def\L{\left[}
\def\R{\right]}
\def\={&=&}
\def\f{\frac}

\def\mc{\mathcal}

\begin{document}
\date{\today }

\title{Preventing the Spread of Online Harms:\\ 
Physics of Contagion across Multi-Platform Social Media and Metaverses}
\author{Chen Xu$^1$, Pak Ming Hui$^2$, Om K. Jha$^3$, Chenkai Xia$^3$, Neil F. Johnson$^3$}
\affiliation{$^{1}$School of Physical Science and Technology, Soochow University, Suzhou 215006, People's Republic of China}
\affiliation{$^{2}$Department of Physics, The Chinese University of Hong Kong, Shatin, Hong Kong SAR, China}
\affiliation{$^{3}$Physics Department, George Washington University, Washington D.C. 20052, U.S.A.}

\begin{abstract}
We present a minimal yet empirically-grounded theory for the spread of online harms (e.g. misinformation, hate) across current multi-platform social media and future Metaverses. New physics emerges from the interplay between the intrinsic heterogeneity among online communities and platforms, their clustering dynamics generated through user-created links and sudden moderator shutdowns, and the contagion process. The theory provides an online `R-nought' criterion to prevent system-wide spreading; it predicts re-entrant spreading phases; it establishes the level of digital vaccination required for online herd immunity; and it can be applied at multiple scales. \end{abstract}
\maketitle

The online spread of scientific misinformation (e.g. concerning COVID-19 vaccines, climate change) and other harms (e.g. hate, racism) is a huge societal problem \cite{Tina,brown,Gill,Cald,DiResta,Larson,threats,rand,gruzd,examplecases,death,gelfand,Qu,usNature2019,usScience2016,Boogs,Brain,Rick,IEEE,Nature2020,Multi2021,30,shapiro1,surge}. The American Physical Society has even launched a 2022 initiative aimed at countering such scientific misinformation \cite{APS}. Preventing online harms from engulfing a Metaverse \cite{metaverse1,metaverse2} will be even more challenging since social media platforms will enhance their existing communication tools with tactile, virtual-reality technologies for remote working, gaming, events, networking and even medical procedures.
There are increasingly impatient calls from governments for social media platforms to do `more' \cite{congress,UK}. But without a clear understanding of how harms spread online at scale, what does `more' mean? 

The structure of the current social media universe and any future Metaverse is complex (Fig. 1(a)). It involves hardware and algorithms of commercially competing platforms (e.g. Facebook,  VKontakte) which then get intertwined by the human activity of $\sim 3$ billion users from across the globe. Their link dynamics are also complex. People aggregate into in-built communities around particular interests (e.g. Facebook `page' or VKontakte `club' for parents) to share advice and experiences \cite{parents1,parents2,parents3}. These in-built communities, many with millions of members, then create links with each other within and across platforms to share content, e.g. a common URL \cite{IEEE,Multi2021}. This facilitates spreading of harmful material at scale (Fig. 1(b)(c)). New intra- and inter-platform links (Fig. 1(a)) can appear daily, and may then get removed en masse during moderator crackdowns on online harms. 

There is a wealth of remarkable studies in physics concerning viral spreading on networks, e.g. by Newman, Barabasi and many others including the cited articles in Refs. \cite{Newman,Barabasi,Menczer,Lamboitte,Holme,Stanley,Vespignani,HH,Gavrilets,Redner,usPRE2010,5,6,7,8,Palla07}. A fascinating example is Watts et al. \cite{Watts}, who showed resurgent epidemics arising in a hierarchical metapopulation model. However we know of no study that combines the features of (1) dynamical clustering of nodes due to link formation and their sudden fragmentation en masse during targeted moderator actions against online harms {\em and} (2) multiple species of node, e.g. 
\begin{figure}[H]
\begin{center}
\epsfig{figure=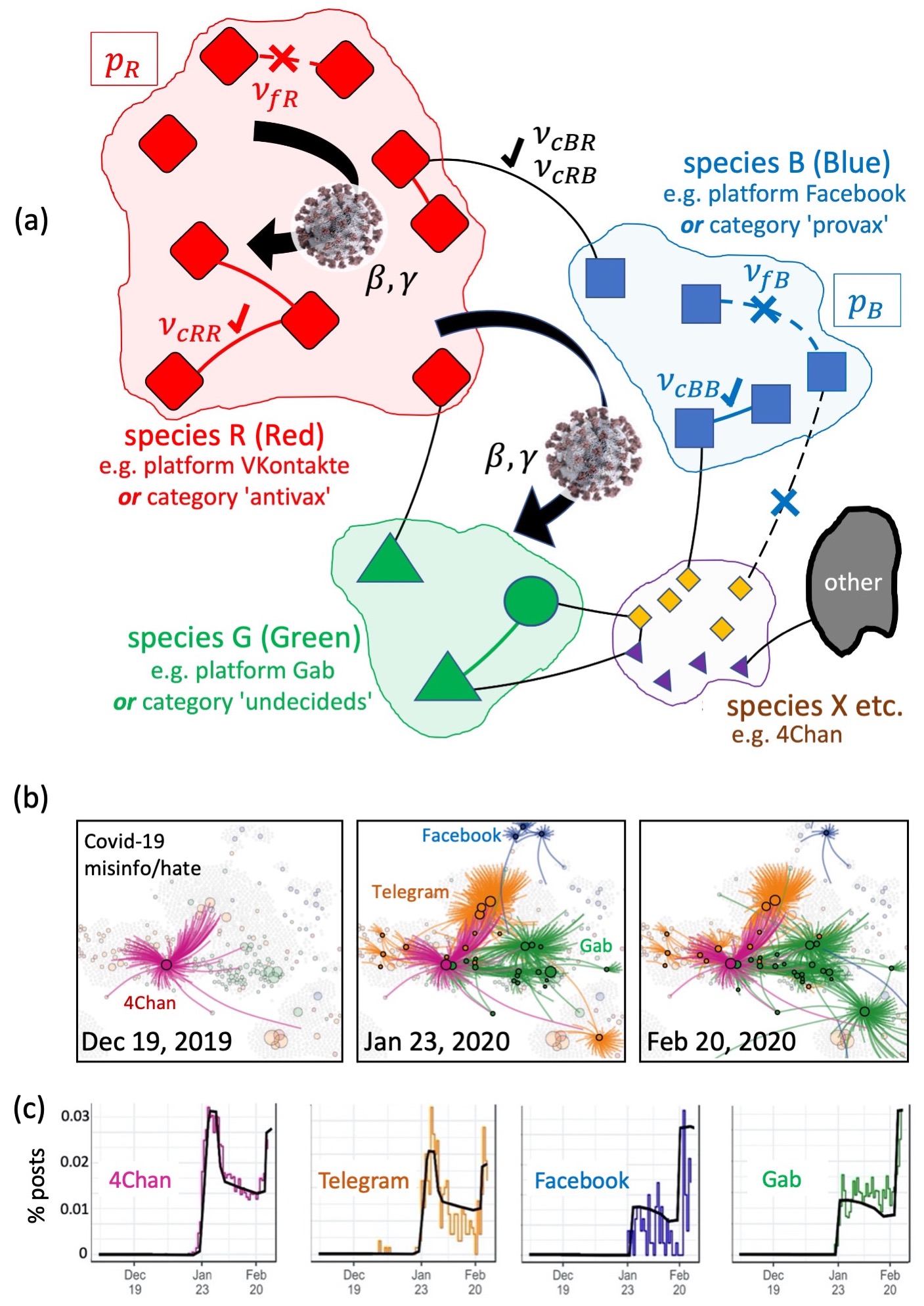,width=1.0\linewidth} 
\caption{(a) Minimal model of online spread of misinformation and other harmful content. Each node is an in-built community such as a Facebook page or VKontakte club. Online links form within and between node species and  get shutdown on masse by moderators, producing node cluster coalescence and fragmentation. (b) Empirical data shows inter- and intra-platform spread of  Covid-19 misinformation/hate during pandemic \cite{Multi2021}. (c) Empirical percentages of harmful material from (b). Sample output from minimal model (black curves) mimics profiles despite all four curves coming from just two model outputs: for 4Chan and Telegram $\gamma=0.01$, for Facebook and Gab $\gamma=0.005$. For all, $v_{cRR}=0.95$, $v_{fR}=0.05$, $\beta=0.05$. Black curves smoothed over timepoints.}
\label{fg:multiverse}
\end{center}
\end{figure}
\noindent in-built communities on different platforms or taking different extreme positions such as anti-vaccination \cite{Nature2020}. Figure 4b of Ref. \cite{IEEE} analyzes in-depth one such cluster involving multiple species of node, which later suddenly fragmented. SM Part B shows more empirical examples.
 
Here we develop a minimal model with these two features, for the spread of online harms (Fig. 1(a)). Our focus is on tipping points for system-wide spreading, not fine-tuning parameters to get the best fits for infection profiles as in Fig. 1(c). 
The nodes, and hence our results, are not limited to in-built communities: they could represent different forms of Multiverse machinery as in Refs. \cite{Cooke,Gorman}. Our results can also be applied at different scales to describe future Metaverses-of-Metaverses etc. by renormalizing the definition of a node and species. 
 
We start with the probability $P_{ij}(t)$ that two nodes $i$ and $j$ selected randomly and independently from across the multi-platform system (platform label $\alpha$ implicit) belong to the same cluster of nodes at time $t$, and hence can share harmful material that they have at time $t$. For  $M=2$ platforms R and B, the Master equation is:
\begin{eqnarray}
\frac{{\rm d} P_{ij}}{{\rm d} t}  =   -P_{ij} \frac{1}{N} \sum_{k \in \{...i...\}}  P_{ki} [\delta_{kR} \nu_{fR} + \delta_{kB} \nu_{fB} ] \ \ \ \ \ \ \ \  \\
 +   [1- P_{ij}]  \f{1}{N} \sum_{m  \in \{...i...\}}  P_{mi} \frac{1}{N} \sum_{n  \notin \{...i...\}}  P_{nj} 
[\delta_{mR}\delta_{nR}\nu_{cRR}\ \ \ \nonumber \\ +\delta_{mB}\delta_{nB}\nu_{cBB}+
\delta_{mR}\delta_{nB}\nu_{cRB}+ \delta_{mB}\delta_{nR}\nu_{cBR}] \nonumber
\end{eqnarray}
\noindent with sums over nodes within a cluster and outside a cluster, and coalescence and fragmentation probabilities per timestep $\{\nu_{c{\alpha}{\alpha'}}\}$ and $\{\nu_{f{\alpha}}\}$. $N$ is the total system node number. These clustering dynamics make sense empirically \cite{IEEE,usNature2019,Multi2021,Nature2020} but can be generalized. Also, the product kernel is consistent with online communications activity data \cite{Palla07,usScience2016} but can be generalized. For $M>2$, the equation is similar, with  platform labels. Averaging Eq. 1 in the steady-state for $M\geq 2$ platforms, and with $p_\alpha$ the fraction of nodes on platform $\alpha$, yields (see SM):
\begin{equation}
 P = {\bar \nu _c}({\bar \nu _c  + N\bar \nu _f })^{-1}
 \end{equation}
\begin{equation}
\bar \nu _c = \sum_{\alpha=1}^M \nu _{c\alpha\alpha} p_\alpha p_\alpha+\sum_{\alpha\neq {\alpha'},1}^M \nu _{c\alpha{\alpha'}} p_\alpha p_{\alpha'},\ \ \ \bar \nu _f  = \sum_{\alpha=1}^M \nu _{f\alpha} p_\alpha
\end{equation}
The SM Fig. 1 shows good agreement between Eq. 2 and simulations.

We adopt a simple ${\mc SIR}$ viral process where $\mc S$, $\mc I$, and $\mc R$ are the numbers of Susceptible nodes (i.e. not yet received misinformation/harmful material {\bf X}), 
Infected nodes (i.e. received {\bf X} and want to share it) and Recovered nodes (i.e. no longer want to share {\bf X}). More complex processes can be treated similarly to what follows, by inserting the appropriate $P$ contributions in their equations.
At each $t$, every $\mc I$ node can infect (i.e. share {\bf X} with) each $\mc S$ node in its cluster with probability $\beta$, and every $\mc I$ node recovers (i.e. becomes an $\mc R$) with probability $\gamma$. 
Empirically, the link dynamics and viral process both evolve on a daily timescale, hence we allow our simulations to have $N$ updates to the link dynamics prior to every update of the $N$-body ${\mc SIR}$ process (see SM for variations of this). When the link-cluster dynamics are in steady-state, we start this viral process by making a Red (R) node infected at $t=0$, e.g. an in-built community on VKontakte posts harmful material {\bf X}. 

We develop the theory at two levels of approximation. Both give reasonable agreement with simulations, but seeing which is closer for a given parameter range can inform the best way to picture the system for that range: 

\noindent (1) Effective Medium Theory (EMT). Here we insert the $P$ from Eq. 2 into the ${\mc SIR}$ equations for the entire population of $\mc S$, $\mc I$, and $\mc R$. Hence the EMT averages within and across species, which crudely speaking tends to underestimate the impact of correlations. This yields ${\dot {\mc S}(t)}=  - \beta P{\mc S}(t){\mc I}(t)$, ${\dot {\mc I}}=  \beta P{\mc S}(t){\mc I}(t) - \gamma {\mc I}$ and ${\dot {\mc R}}=  \gamma {\mc I}$. Setting ${\dot {\mc I}}>0$ gives a generalized `R-nought' spreading condition ${\kappa}^{\rm EMT}=(N-1)P\beta /\gamma>1$, i.e. $N P\beta /\gamma >1$ for large $N$. Hence an approximate analytic condition for preventing system-wide spreading is: 
\begin{equation}
{\kappa}^{\rm EMT}\equiv (N-1) {\bar \nu _c} \beta [({\bar \nu _c  + N\bar \nu _f })\gamma]^{-1} < 1\ .
\end{equation}
\noindent Equation 4 shows explicitly the interplay, and hence trade-offs for intervention, between the clustering dynamics (${\bar \nu _c}$ and ${\bar \nu _f}$) and the viral dynamics ($\beta$ and $\gamma$), or equivalently their respective timescales given by the reciprocals. A key implication is that even if $\beta/\gamma > 1$, which means system-wide spreading is predicted using conventional ${\mc SIR}$ equations, the no-spreading condition in Eq. 4 can still be satisfied if ${\bar \nu _c}$ is sufficiently small and/or
${\bar \nu _f}$ is sufficiently large. In short, {\em the link dynamics can prevent system-wide spreading even though the implicit contagiousness of the material is sufficiently high} ($\beta/\gamma>1$). The fact that so much material is spreading system-wide online, suggests that current ${\bar \nu _c}$ and ${\bar \nu _f}$ values need adjusting. This finding also raises questions about mitigation strategies that focus exclusively on the implicit contagiousness ($\beta/\gamma$) of a specific piece of material, as measured in some offline experiment.
\begin{figure}
\epsfig{figure=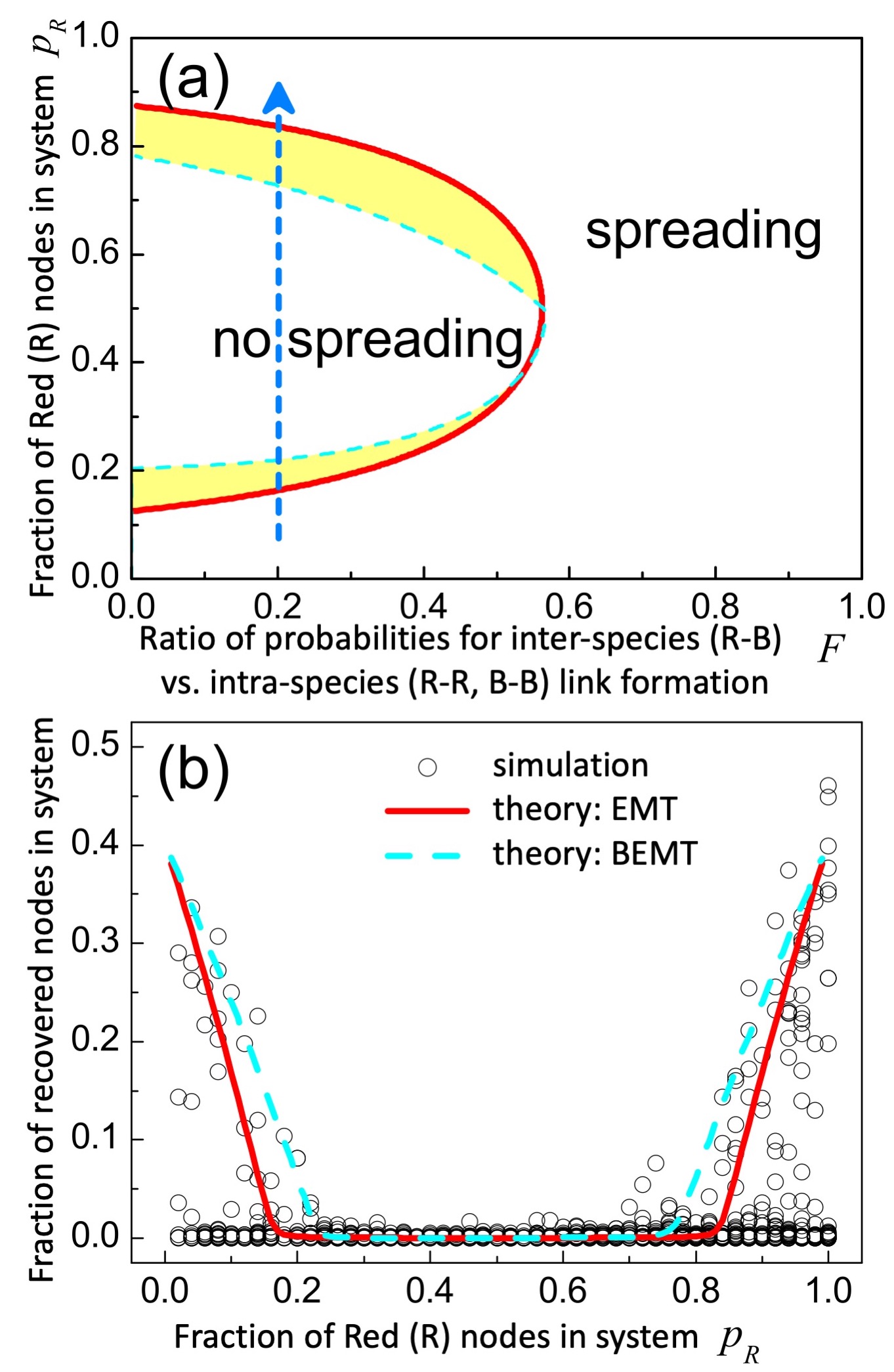,width=0.85 \linewidth}
\caption{Impact on system-wide spreading when probability of inter-species (e.g. inter-platform) link formation is a fraction $F<1$ of probability of intra-species (e.g. intra-platform) link formation (i.e. Case I). (a) Phase diagram for system-wide spreading. In Figs. 2-4, red (blue) curve is EMT (BEMT) theory which tends to underestimate (overestimate) correlations, hence yellow area is crude indicator of spread in simulation output (see SM for agreement). $\nu_f=0.025$, $\nu_c=0.32$, $\beta=0.005$, $\gamma=0.05$. In Figs. 2-4, $N=10000$ and one R node is infected at $t=0$ with link dynamics already in steady state. (b) Simulation and theory for $F=0.2$ and increasing $p_R$, which corresponds to dotted vertical line in (a).}  \label{fg:Figure2}
\end{figure}

\noindent (2) Beyond Effective Medium Theory (BEMT). Here we insert the species-specific contributions within $P$ (see Eq. 3) into the coupled ${\mc SIR}$ equations for R and B separately (see SM Eqs. 27-32). Hence the BEMT only averages within a species, which crudely speaking tends to overestimate the impact of correlations. The approximate analytic condition for preventing initial system-wide spreading is  (see SM Eq. 37 for derivation):
  
{\small\begin{equation}
{\kappa}^{\rm BEMT}\equiv \frac{{(N-1) P}}{{p_R \bar \nu _c}}
\big[\nu _{cRR} p_R^2 +\frac{1}{2}(\nu _{cRB}  + \nu _{cBR} )p_B p_R
\big]\frac{\beta}{\gamma} < 1
\end{equation}}
\noindent Equation 5 also predicts that system-wide spreading can be prevented even if $\beta/\gamma>1$. 
The EMT and BEMT also both mirror the full simulations in predicting that although viral material can appear isolated and largely eradicated on a given platform, it can simultaneously be moving through inter-cluster links to other platforms where it revives before later re-emerging on the original platform. This means that  moderators reviewing Facebook (blue) clusters in January 2020 (Fig. 1(b)) might conclude they are getting rid of Covid-19 misinformation, only to see it later re-emerge via other platforms. This in turn suggests system-wide spreading of harmful content cannot be controlled by a single platform.

\begin{figure}
\begin{center}
\epsfig{figure=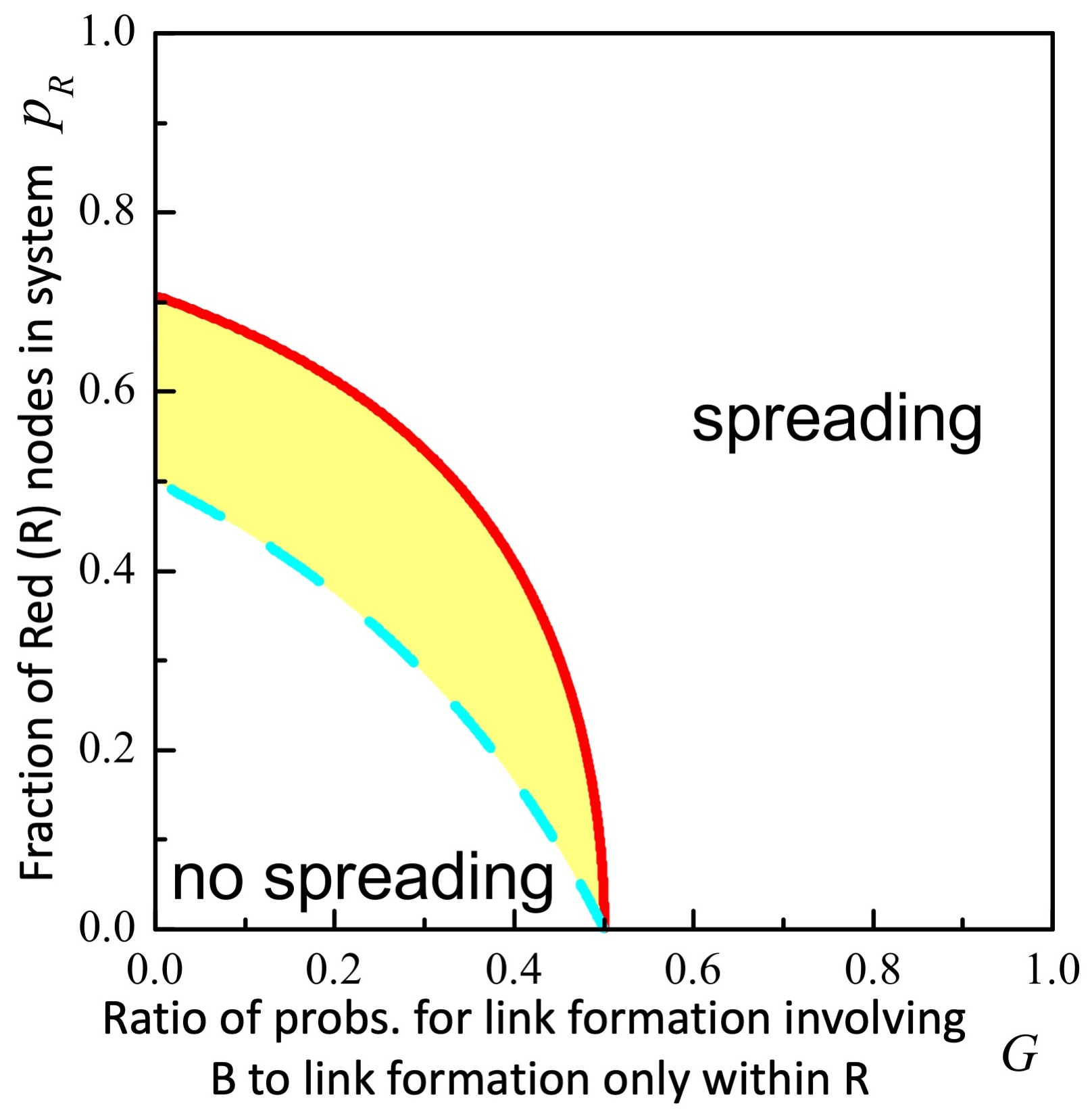,width=0.65\linewidth}
\caption{Impact on system-wide spreading when probability of link formation involving platform B is a fraction $G<1$ of the probability of link formation within R only (i.e. Case II). Phase diagram for system-wide spreading.  $\nu_f=0.05$, $\nu_c=0.5$, $\beta=0.01$, $\gamma=0.05$. 
}  \label{fg:Figure3}
\end{center}
\end{figure}
\begin{figure}
\begin{center}
\epsfig{figure=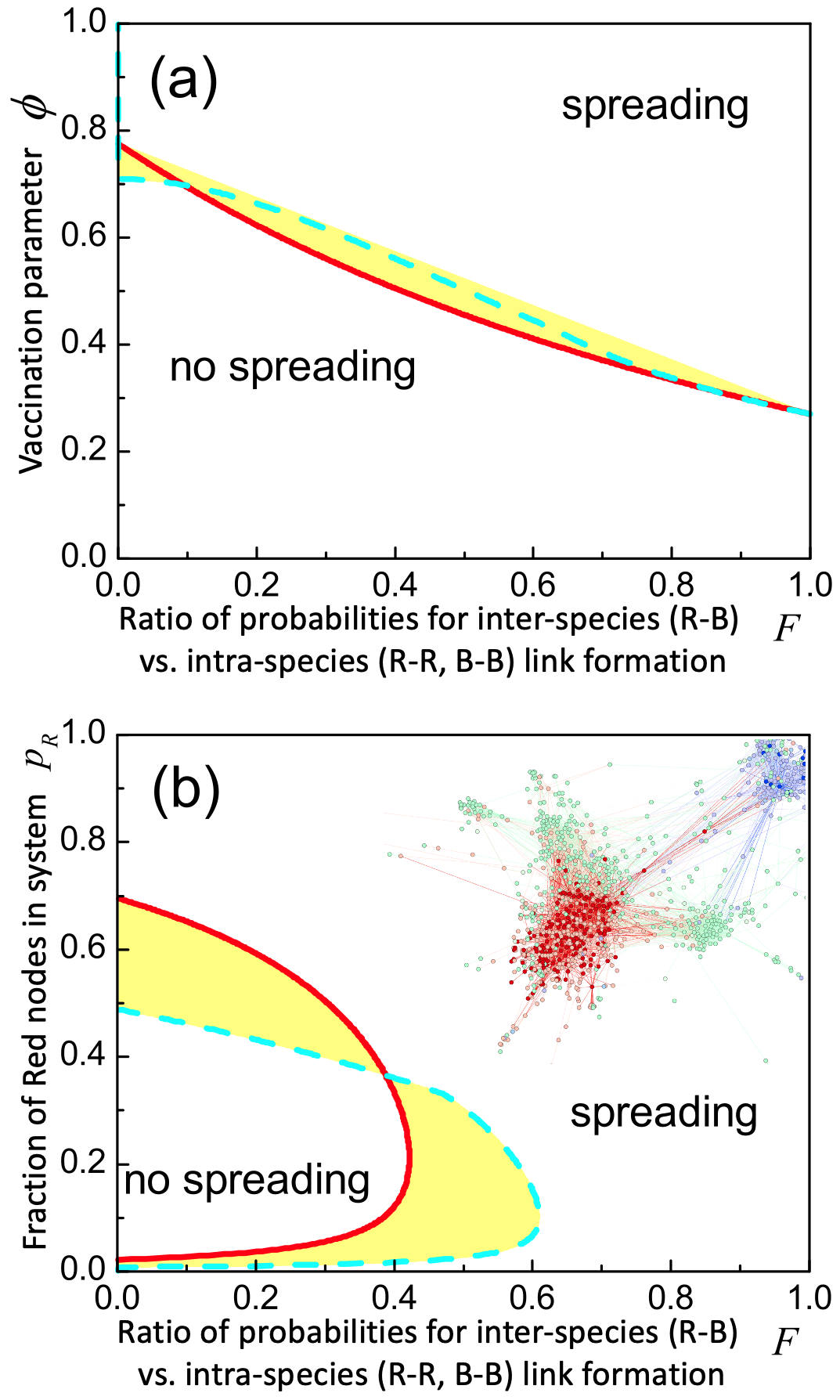,width=0.85\linewidth}
\caption{Impact of online vaccination on system-wide spreading. Here the B nodes are the vaccinated subset of all $N$ nodes: their infection rate is $\chi=\phi\beta<\beta$. (a) Boundary shows the $\phi$ values required to achieve no system-wide spreading for a given $F$, i.e. to achieve online herd immunity. Link probabilities as in Case I, with $p_R=0.3$. (b) Vaccination distorts phase diagram compared to Fig. 2(a). $\phi=0.5$, $\nu_f=0.025$, $\nu_c=0.32$, $\beta=0.008$, $\gamma=0.05$. Inset shows 
Facebook pages (each is a node) in December 2020 surrounding the vaccine debate. Red (blue) nodes are anti (pro) vaccination (see Ref. \cite{IEEE}). Green nodes are neutral. Darker nodes of each color are pages (nodes) that are vaccinated, i.e. they feature the Facebook message promoting best-science vaccine guidance. Their total fraction is $p_B$. Lighter nodes of each color are not vaccinated (i.e. total fraction $p_R=1-p_B$).  Most vaccinated nodes are antis (red) but not all reds are vaccinated.} \label{fg:Figure4}
\end{center}
\end{figure}
First, we apply our theory to the empirically relevant scenario where the probability of {\em inter}-species (e.g. inter-platform) link appearances differs from {\em intra}-species (e.g. intra-platform) link appearances (Case I) as observed empirically for VKontakte (R) and Facebook (B) \cite{Multi2021}. We set $\nu_{cRB}=\nu _{cBR} = F \nu_c$ with 
$\nu_{cRR}=\nu_{cBB}=\nu_c$, and for simplicity $\nu_{fR}=\nu_{fB}=\nu_f$. $0\leq F < 1$ means the probability of link formation is smaller for nodes from different species (e.g. platforms) than for the same species (e.g. platform).
Figure 2 shows the theoretical phase boundaries and illustrates the good agreement with simulation. For this specific parameter range, the simpler EMT appears  sufficient, though the SM confirms that the BEMT is better elsewhere in the parameter space. 
As expected, both theories predict that R (e.g. VKontakte) can prevent spreading from it by reducing $F$ -- but non-trivially this is {\em only} true if both R and B have comparable fractions of nodes ($p_R$ not too far from $0.5$). Specifically, if (and only if) $p_R$ falls within the semi-ellipse shown, $F$ can be finite and  yet still system-wide spreading will not occur -- this allows platforms to avoid blanket `social distancing' or `quarantining' from each other, and also avoid claims of over-policing. Figure 2(a) 
also suggests that with $F$ fixed and small, and $p_R$ evolving so slowly that the theories remain valid (i.e. clustering-viral dynamics much faster than changes in $p_R$), the system-wide spreading will exhibit re-entrant phases (vertical blue line). This warns that if a platform R finds itself moving into a no-spreading regime as its popularity over platform B grows (i.e. $p_R$ increases from a small initial value), it cannot assume it is out of danger since it can suddenly tip to a spreading regime as $p_R$ gets larger.

Second, we apply our theory to the empirically relevant scenario of {\em platform-dependent} (or more generally, species dependent) frequency of link appearances (Case II) as observed for example when comparing Facebook and Gab due to Facebook's far higher levels of activity \cite{Multi2021}.
We mimic this by setting $\nu_{cRB}=\nu _{cBR}=\nu_{cBB}=G\nu_c$ with
$\nu_{cRR}=\nu_c$. $0\leq G \leq 1$ means B nodes (e.g. nodes on platform B) are less likely to form links as compared to R nodes (e.g. nodes on platform R). Figure 3 shows the theoretical phase boundaries: 
no-spreading requires small $G$ and small $p_R$, which makes a dominant (i.e. large $p_R$) R platform less likely to be able to prevent multi-platform spreading irrespective of how it controls its frequency of link appearances relative to other platforms.

Third, we use our theory to quantify the online herd immunity required to prevent system-wide spreading. The concept of a `digital vaccination' is already being pursued by Facebook: it posts positive counter-messages on Facebook pages (nodes) to try to prevent misinformation taking hold (see inset Fig. 4(b) \cite{IEEE,Nature2020}). Vaccinated nodes  have infection probability $\beta$ reduced to $\chi=\phi\beta$ where $0\leq\phi<1$. So ${\dot {\mc S}_R(t)}  = -\beta P {\mc S}_R(t) {\mc I}(t)$, ${\dot {\mc S}_V(t)}  = -\chi P {\mc S}_V(t) {\mc I}(t)$,  ${\dot {\mc I}}(t)  = P (\beta {\mc S}_R(t)  + \chi {\mc S}_V(t) ){\mc I}(t) - \gamma {\mc I}(t)$ within EMT ($V\equiv B$ hence $p_B\equiv p_V$ etc.).
Hence the critical fraction of all $N$ nodes that need to be vaccinated to prevent system-wide spreading, is:
\begin{equation}
p_V \geq \f{1}{1-\phi} \L 1-\f{\gamma}{PN\beta}\R \ . 
\end{equation}
\noindent Equation 6 also applies when it is only the nodes on a given platform B that are all vaccinated, i.e. $p_B\equiv p_V$ is fixed, in which case Eq. 6 predicts the level of vaccine efficacy $(1-\phi)$ required to prevent system-wide spreading. Figure 4(a) shows the boundary given by solving for $\phi$. The shift to higher $F$ as $\phi$ decreases means that increasing vaccine efficacy $(1-\phi)$ allows higher probabilities of inter-species (i.e. inter-platform) link formation $F$ to be tolerated while still avoiding system-wide spreading. Figure 4(b) is similar to Fig. 2(a), but shows the impact of having all B nodes vaccinated, i.e. the phase boundaries are distorted but the size of the spreading region does not change much. This provides a warning: the inter-platform link dynamics play such a crucial role that vaccination of one species or platform can simply end up refocusing the harmful material toward another. This was observed during the pandemic, where action against misinformation among the anti-vaccination nodes on Facebook helped push it to other platforms where it then proliferated.

Much remains to be done, e.g. there are many types of platform and harmful material $\bf X$, and the `infection' terminology is imperfect. But our mathematical results can be easily generalized to multiple platforms, and to allow different $\bf X$ to have differing parameter values (e.g. $\beta$ and $\gamma$). Our results can also be applied at different levels of granularity, e.g. different categories within or across platforms with R (B) being anti- (pro-) establishment science in association with climate change debates.

NFJ acknowledges support of AFOSR grants FA9550-20-1-0382 and FA9550-20-1-0383. We are grateful to Nicholas J. Restrepo, Rhys Leahy, Nicolas Velasquez and Yonatan Lupu for help in obtaining the empirical data in Fig. 1, and King Yan Fong for additional help.


\begin{thebibliography}{99}
\bibitem{Tina} N. Calleja et al. 
A public health research agenda for managing infodemics: Methods and results of the first WHO infodemiology conference. JMIR Infodemiology 1, e30979 (2021). \url{DOI: 10.2196/30979}
\bibitem{brown}	 R. Brown et al. Counteracting Dangerous Narratives in the Time of COVID-19. Over Zero (2020). \url{https://projectoverzero.org/newsandpublications}
\bibitem{Gill} P. Gill, E. Corner. Lone-actor terrorist use of the Internet and behavioral correlates. In Terrorism Online: Politics, Law, Technology and Unconventional Violence. Eds. L. Jarvis, S. Macdonald, T.M. Chen (Routledge, London, 2015)
\bibitem{Cald} A. Bessi, M. Coletto, G. A. Davidescu, A. Scala, G. Caldarelli, W. Quattrociocchi. Science vs conspiracy: Collective narratives in the age of misinformation. PLoS One 10, 0118093, (2015). 
\bibitem{DiResta} R. DiResta. Of virality and viruses: the anti-vaccine movement and social media. NAPSNet Special Reports, 08-Nov-2018. 
\bibitem{Larson} H. Larson. A lack of information can become misinformation. Nature 580, 306 (2020)
\bibitem{threats} B. Nogrady. I hope you die: how the COVID pandemic unleashed attacks on scientists. Nature 598, 250 (2021)
\bibitem{rand} G. Pennycook, D. G. Rand. Who falls for fake news? The roles of receptivity, overclaiming, familiarity, and analytic thinking. Journal of Personality 88, 185 (2019)
\bibitem{gruzd}	A. Gruzd, P. Mai. Inoculating Against an Infodemic: A Canada-Wide COVID-19 News, Social Media, and Misinformation Survey. SSRN Electronic Journal, 2020. \url{DOI: http://dx.doi.org/10.2139/ssrn.3597462} 
\bibitem{examplecases} For examples of threats against celebrities, see \url{https://www.bbc.com/sport/football/56714760}
\bibitem{death} BBC News. Covid-19: French MPs get death threats over support for vaccine pass. Jan 3, 2022. \url{https://www.bbc.com/news/world-europe-59860058}
\bibitem{gelfand} M. J. Gelfand, J. R. Harrington, J. C. Jackson. The Strength of Social Norms Across Human Groups. Perspectives on Psychological Science: Journal of the Association for Psychological Science 12, 800 (2017). \url{DOI: 10.1177/1745691617708631}
\bibitem{Qu} M. Cinelli, W. Quattrociocchi, A. Galeazzi, C.M. Valensise, E. Brugnoli, A.L. Schmidt, P. Zola, F. Zollo, A. Scala. The COVID-19 social media infodemic. Sci. Rep. 10, 16598 (2020)
\bibitem{usNature2019} N.F. Johnson, R. Leahy, N. Johnson Restrepo, N. Velasquez, M. Zheng, P. Manrique, P. and S. Wuchty. Hidden resilience and adaptive dynamics of the global online hate ecology. Nature 573, 261 (2019)
\bibitem{usScience2016} N.F. Johnson, M. Zheng, Y. Vorobyeva, A. Gabriel, N. Velasquez, P. Manrique, D. Johnson, E. Restrepo, C. Song and S. Wuchty. New online ecology of adversarial aggregates: ISIS and beyond. Science 352, 1459 (2016)
\bibitem{Boogs} N. Velasquez, P. Manrique, R. Sear, R. Leahy, N. Johnson Restrepo, L. Illari, Y. Lupu, N.  F. Johnson. Hidden order in online extremism and its disruption by nudging collective chemistry. Sci. Rep. 11, 9965 (2021)
\bibitem{Brain} M. Zheng, Z. Cao, Y. Vorobyeva, P. Manrique, C. Song,
N.F. Johnson. Multiscale dynamical network mechanisms underlying aging from birth to death. Sci. Rep. 8, 3552, (2018)
\bibitem{Rick} R. F. Sear, N. Velasquez, R. Leahy, N. Johnson Restrepo, S. El Oud, N. Gabriel, Y. Lupu, N. F. Johnson. Quantifying COVID-19 Content in the Online Health Opinion War Using Machine Learning. IEEE Access 8, 91886 (2020). \url{DOI: 10.1109/ACCESS.2020.2993967}
\bibitem{IEEE} N. J. Restrepo, L. Illari, R. Leahy, R. F. Sear, Y. Lupu and N. F. Johnson. How Social Media Machinery Pulled Mainstream Parenting Communities Closer to Extremes and their Misinformation during Covid-19. IEEE Access \url{DOI: 10.1109/ACCESS.2021.3138982} (2022)
\bibitem{Nature2020} N.F. Johnson, N. Velasquez, N. Johnson Restrepo, R. Leahy, N. Gabriel, S. El Oud, M. Zheng, P. Manrique, S. Wuchty, Y. Lupu. The online competition between pro- and anti-vaccination views. Nature 582, 230 (2020). \url{DOI: https://doi.org/10.1038/s41586 020-2281-1}
\bibitem{Multi2021} N. Velasquez, R. Leahy, N. Johnson Restrepo, Y. Lupu, R. Sear, N. Gabriel, O. K. Jha, B. Goldberg, N. F. Johnson. Online hate network spreads malicious COVID-19 content outside the control of individual social media platforms. Sci. Rep. 11, 11549 (2021)
\bibitem{30} I. Van Der Vegt, M. Mozes, P. Gill, B. Kleinberg. Online influence, offline violence: Linguistic responses to the `Unite the Right' rally (2019). \url{https://arxiv.org/ftp/arxiv/papers/1908/1908.11599.pdf}
\bibitem{shapiro1} A. Redi Ross, M. Modi, P. Paresky, L. Jussim, A. Goldenberg, P. Goldenberg, D. Finkelstein, J. Farmer, K. Holden, D. Riggleman, J. Shapiro, J. Finkelstein. Network Contagion Research Institute report. March 11, 2021. A contagion of institutional distrust: Viral Disinformation of the COVID Vaccine and the Road to Reconciliation. \url{https://networkcontagion.us/reports/a-contagion-of-institutional-distrust/}
\bibitem{surge} S. Frenkel, D. Alba, R. Zhong. Surge of Virus Misinformation Stumps Facebook and Twitter. The New York Times, 08-Mar-2020. \url{ www.nytimes.com/2020/03/08/technology/coronavirus- misinformation-social-media.html}
\bibitem{APS} APS Set to Launch New Initiative to Oppose Scientific Misinformation This January. Available at \url{https://engage.aps.org/fps/resources/newsletters/january-2022#misinfo}
\bibitem{metaverse1} S. Frenkel, K. Browning. The Metaverse’s Dark Side: Here Come Harassment and Assaults. The New York Times. Dec 30, 2021. \url{https://www.nytimes.com/2021/12/30/technology/metaverse-harassment-assaults.html}
\bibitem{metaverse2} P.A. Clark. The Metaverse Has Already Arrived. Here’s What That Actually Means. Time. Nov 15, 2021. \url{https://time.com/6116826/what-is-the-metaverse/}
\bibitem{congress} B. Fung. Facebook, Twitter and Google CEOs grilled by Congress on misinformation. CNN March 25, 2021. \url{https://www.cnn.com/2021/03/25/tech/tech-ceos-hearing/index.html}
\bibitem{UK} The UK Home Affairs Select Committee. Hate Crime: Abuse, Hate and Extremism Online. Session 2016–17 HC 609. \url{https://publications.parliament.uk/pa/cm201617/cmselect/cmhaff/609/609.pdf}
\bibitem{parents1} R.Y. Moon, A. Mathews, R. Oden, R. Carlin. Mothers’ Perceptions of the Internet and Social Media as Sources of Parenting and Health Information: Qualitative Study. Journal of Medical Internet Research 21, 14289 (2019)
\bibitem{parents2}  T. Ammari, S. Schoenebeck. Thanks for Your Interest in Our Facebook Group, but It's Only for Dads: Social Roles of Stay-at-Home Dads. In CSCW '16, San Francisco, CA, USA, Feb 27-March 2, 2016. 
\bibitem{parents3} R. Laws, A.D. Walsh, K. D. Hesketh, K.L. Downing, K. Kuswara, K. J. Campbell. Differences Between Mothers and Fathers of Young Children in Their Use of the Internet to Support Healthy Family Lifestyle Behaviors: Cross-Sectional Study. Journal of Medical Internet Research 21, 11454 (2019)
\bibitem{Newman} M.E. Newman. Networks (Oxford University Press, 2016)
\bibitem{Barabasi} A-L Barabasi. Network Science (Cambridge University Press, 2016)
\bibitem{Menczer} F. Menczer, S. Fortunato. A First Course in Network Science (Cambridge University Press, 2020)
\bibitem{Lamboitte} N. Masuda, R. Lamboitte. Guide to Temporal Networks (World Scientific Publishing, 2020)
\bibitem{Holme} N. Masuda, P. Holme. Temporal Network Epidemiology (Springer, 2018)
\bibitem{Stanley} X. Chen, R. Wang, M. Tang, S. Chai, H.E. Stanley, L.A. Braunstein. Suppressing epidemic spreading in multiplex networks with social-support.
New J. Phys. 20, 013007 (2018)
\bibitem{Vespignani} R. Pastor-Satorras, A. Vespignani. Epidemic Spreading in Scale-Free Networks. Phys. Rev. Lett. 86, 3200 (2001)
\bibitem{HH} A. Soulier, T. Halpin-Healy. The Dynamics of Multidi- mensional Secession: Fixed Points and Ideological Con- densation. Phys. Rev. Lett. 90, 258103 (2003)
\bibitem{Gavrilets} S. Gavrilets. Collective action and the collaborative brain. J. R. Soc. Interface 12, 20141067 (2015)
\bibitem{Redner} P.L. Krapivsky, S. Redner and E. Ben-Naim. A Kinetic View of Statistical Physics (Cambridge University Press, Cambridge, 2010)
\bibitem{usPRE2010} Z. Zhao, J.P. Calderon, C. Xu, G. Zhao, D. Fenn, D. Sornette, R. Crane, P.M. Hui, N.F. Johnson. Effect of social group dynamics on contagion. Phys. Rev. E 81, 056107 (2010)
\bibitem{5} V. Colizza, A. Barrat, M. Barthlemy, A. Vespignani. The role of the airline transportation network in the prediction and predictability of global epidemics. Proc. Natl. Acad. Sci. USA 103 2015 (2006)
\bibitem{6} J.T. Davis, N. Perra, Q. Zhang, Y. Moreno, A. Vespignani. Phase transitions in information spreading on structured populations. Nat. Phys. 16, 590 (2020)\bibitem{7} W. Quattrociocchi, G. Caldarelli, A. Scala. Opinion dynamics on interacting networks: media competition and social influence. Sci Rep 4, 4938 (2014)
\bibitem{8} J.P. Onnela, J. Saramaki, J. Hyvonen, G. Szabo, D. Lazer, K. Kaski, J. Kertesz, A. L. Barabasi. Structure and tie strengths in mobile communication networks. PNAS 104, 7332 (2007)
\bibitem{Palla07} G. Palla, A., Barabasi, T. Vicsek. Quantifying social group evolution. Nature 446, 664 (2007)
\bibitem{Watts} D.J. Watts, R. Muhamad, D.C. Medina, P.S. Dodds. Multiscale, resurgent epidemics in a hierarchical metapopulation model. Proc. Natl. Acad. Sci. USA 102, 11157 (2005)
\bibitem{Cooke} J.C. Gorman, M. Demir, N.J. Cooke, D.A. Grimm. Evaluating sociotechnical dynamics in a simulated remotely-piloted aircraft system: a layered dynamics approach. Ergonomics 62, 629 (2019). \url{https://doi.org/10.1080/00140139.2018.1557750}
\bibitem{Gorman} J.C. Gorman, D.A. Grimm, R.H. Stevens, T. Galloway, A.M. Willemsen-Dunlap, D.J. Halpin. Human Factors 62, 825 (2020). \url{DOI: 10.1177/0018720819852791}



\end{thebibliography}
\end{document}